\documentclass[a4paper,fleqn,usenatbib]{mnras}

\usepackage[T1]{fontenc}
\usepackage{ae,aecompl}

\usepackage{graphicx}	% Including figure files
\usepackage{amsmath}	% Advanced maths commands
\usepackage{amssymb}	% Extra maths symbols
\usepackage{subfig}     % Including subfigures
\usepackage{breakurl}   % Line-breakable url-like links in hyperref 
\usepackage{float}
\usepackage[section]{placeins}
\title[Profiles and timing of PSR J1757$-$2421]{Pulse profiles and timing of PSR J1757$-$2421}

\author[J. P. Yuan et al.]{
 J. P. Yuan,$^{1,2}$\thanks{E-mail: yuanjp@xao.ac.cn}
 R. N. Manchester,$^{3}$
 N. Wang,$^{1,2}$
 J. B. Wang,$^{1}$
X. Zhou,$^{1}$
\newauthor W. M. Yan,$^{1}$ 
and Z. Y. Liu,$^{1}$
\\
 $^{1}$Xinjiang Astronomical Observatory, CAS,
  150 Science 1-Street, Urumqi, Xinjiang, China, 830011\\
$^{2}$Key Laboratory of Radio Astronomy, Chinese Academy of Science, Nanjing,
 China, 210008 \\
$^3$CSIRO Astronomy and Space Science, Australia Telescope National Facility, 
 PO Box 76, Epping, NSW 1710, Australia \\
}

\date{Accepted ... Received ...; in original form 2016 Sept.}

\pubyear{2016}

\begin{document}
\label{firstpage}
\pagerange{\pageref{firstpage}--\pageref{lastpage}}
\maketitle

\begin{abstract}

Pulse arrival time measurements over fourteen years with the Nanshan
25-m and Parkes 64-m radio telescopes have been used to determine the
average profile and timing properties for PSR J1757$-$2421
(B1754$-$24). Analysis of the radio profile data shows a large
variation of spectral index across the profile and an unusual increase
in component separation with increasing frequency. The timing
observations show that PSR B1754$-$24 underwent a large glitch with a fractional
increase in spin frequency of  $\Delta\nu_{\rm g}/\nu\sim 7.8\times10^{-6}$ in
May 2011. At the time of the glitch there was a large permanent jump
in the rotational frequency accompanied by two exponential recovery
terms with timescales of 15(6) and 97(15) days, respectively.
\end{abstract}


%%%%%%%%%%%%%%%%%%%%%%%%%%%%%%%%%%%%%%%%%%%%%%%%%%

%%%%%%%%%%%%%%%%% BODY OF PAPER %%%%%%%%%%%%%%%%%%
%\FloatBarrier
\section{Introduction} 

Pulsars are rapidly rotating and strongly magnetised neutron stars
that radiate beams tied to the rotating star through their magnetic
fields. Frequent timing observations over months and years can give an
accurate model for the spin behaviour of the neutron star. Although
pulsars are generally precise clocks, the timing observations have
detected the two types of instability of pulsar rotation, glitches and
timing noise. A glitch is characterised by abrupt increase in the
rotation frequency of pulsar, usually accompanied by an increase in
the spin-down rate.  This event is often followed by an exponential
recovery towards the extrapolation of the pre-glitch behaviour.  Up to
now, more than 400 glitch events have been detected in at least 140
pulsars, with more than 200 having been published since
2010\footnote
  {\url{http://www.atnf.csiro.au/people/pulsar/psrcat/}(version 1.54,
    \citealt{mhth05});
    \url{http://www.jb.man.ac.uk/pulsar/glitches.html}}. Observed
fractional glitch sizes range from $\sim10^{-10}$ to $\sim10^{-5}$
with a bimodal distribution, where the peaks lie around
$2\times10^{-9}$ and $10^{-6}$ respectively
\citep{ywml10,elsk11,ymh+13}.

Since the phenomenon was first detected in the Vela pulsar
\citep{rm69}, there has been a lot of effort to uncover the glitch
mechanism, nevertheless, it is not fully understood.  The early
studies suggested that glitches result from crustquakes in the neutron
star \citep[e.g.,][]{bp71}.  When the starquake model proved
unviable, it was suggested that glitches are triggered by decoupling
and recoupling of vortices between the crust and inner superfluid, resulting in the
transfer of angular momentum to the crust \citep[e.g.,][]{ai75,aaps81}. 
  \citet{ga09} suggested that an r-mode instability may trigger a
  global unpinning of vortices leading to a glitch.  The vortex
  unpinning model can explain the various aspects of large glitches
  and post-glitch behaviour \citep[e.g.,][]{ha14}, but \citet{zltx14} have
  revived the starquake model, suggesting that material is transported
  from the equator to the pole, causing the star spin up, with large
  jumps of $\dot\nu/\nu\sim10^{-6}$ possible. 

\begin{table*}%[!htb]
\centering
\caption{Details of observations for PSR J1757$-$2421}.
\label{tb1757ob}
\begin{tabular}{llcrlrrl}
\hline
Telescope & Frontend & Central freq. & Bandwidth & Backend   &  Pulse phase bins &  ToAs & Data span\\
          &          & (MHz)             & (MHz)     &         &             &  \\
\hline
Nanshan 25 m & Room-temp.       & 1540 &  320 $~$ & $~$AFB   &  256  $~$  $~$ & 110$~$  & Jan 06, 2000 -- Jul 25, 2002 \\
Nanshan 25 m & Cryogenic        & 1540 &  320 $~$ & $~$AFB   &  256  $~$  $~$ & 280$~$  & Jul 29, 2002 -- Mar 04, 2014 \\
Nanshan 25 m & Cryogenic        & 1540 &  320 $~$ & $~$DFB   &  512  $~$  $~$ &  69$~$  & Jan 18, 2010 -- Mar 04, 2014 \\
%\hline                                                                          
Parkes  64 m & Multibeam        & 1374 &  288 $~$ & $~$AFB   &  512  $~$  $~$ &  20$~$  & Dec 01, 2000 -- Sep 30, 2006 \\
Parkes  64 m & Multibeam        & 1369 &  256 $~$ & $~$PDFB3 & 1024  $~$  $~$ &  56$~$  & Jul 24, 2008 -- Feb 04, 2013 \\
Parkes  64 m & Multibeam        & 1369 &  256 $~$ & $~$PDFB4 & 1024  $~$  $~$ &   8$~$  & Feb 19, 2009 -- Nov 10, 2009 \\
Parkes  64 m & 1050cm           & 3094 & 1024 $~$ & $~$PDFB2 & 1024  $~$  $~$ &   5$~$  & Jul 23, 2007 -- Jul 23, 2008 \\
Parkes  64 m & 1050cm           & 3094 & 1024 $~$ & $~$PDFB4 & 1024  $~$  $~$ &   9$~$  & Jan 01, 2009 -- Jan 18, 2013 \\
\hline
\end{tabular}
\end{table*}

 Timing noise is a continuous, erratic fluctuation of the
  observed pulsar period, usually modelled as a red noise process in
  the timing residuals with a correlation on time-scales of months to
  years. Timing noise may result from a random walk in the pulse
  phase, spin or spin down, which would give power-law spectra
  for the timing residuals ($S(f)\propto f^{\alpha}$) with spectral
  indices $\alpha=-2,-4$ or $-6$ respectively \citep{bgh+72}.  But there
is some argument that it cannot simply be modelled as a
random walk \citep[e.g.,][]{cd85,hlk10}.  It may be generated by
various intrinsic sources, such as superfluid turbulence
\citep{gre70,ml14}, microglitches \citep{cd85,mpw08}, variable
coupling between the crust and liquid interior or pinned and
corotating regions \citep{anp86,jon90}, fluctuations in the spin-down
torque \citep{che87,ulw06}, post-glitch recovery \citep{jg99},
magnetospheric state switching \citep[e.g.,][]{lyn10}, asteroid belts
 \citep{cs08}, evolution of the magnetic inclination angle
\citep{yz15} and fluctuations in the magnetosphere \citep{otkd16}.

PSR J1757$-$2421 (B1754$-$24) was discovered during a 1974 pulsar
survey with Parkes 64m radio telescope \citep{kom74}. This star has a
spin period $P=0.234$~s and period derivative $\dot P =
1.29\times10^{-14}$ s s$^{-1}$.  It has a characteristic age
$\tau_{\rm c} = P/(2\dot P) = 2.87\times10^5$ yr and a surface
  dipole magnetic field strength of $1.76\times10^{12}$
  G. \citet{hllk05} obtained a proper motion of PSR J1757$-$2421 in
  ecliptic longitude of $-17(5)~\rm{mas~yr}^{-1}$ from pulse timing
  observations.  Even though PSR J1757$-$2421 is energetic with a
  spin-down power ($\dot E$) of $4.0\times10^{34}~ \rm{erg~s}^{-1}$
  and it is on the list of gamma-ray pulsar candidates for the Fermi
  Telescope \citep{sgc+08}, there is no report of detection of
  gamma-ray emission from this pulsar\footnote{\url{
 https://confluence.slac.stanford.edu/display/GLAMCOG/Public+List+of+LAT-
 Detected+Gamma-Ray+Pulsars}}.

In this article, we report on the pulsar profiles and timing behaviour for PSR 
J1757$-$2421, especially the large glitch which occurred in 2011.

\section{Observations}
In the pulsar timing program carried out with the Nanshan 25-m radio
telescope located in Xinjiang, China, observations were taken with a
receiver which has a bandwidth of 320 MHz centred at 1540 MHz.  The
monitoring of PSR J1757$-$2421 was conducted with an observing cadence
of three times per month and a 16-min duration.  Early observations
prior to 2010 were taken using an analogue filterbank (AFB) with
128$\times$2.5 MHz subchannels and are described in \citet{wmz+01}.
After January 1, 2010, data were also obtained with a digital filterbank
(DFB) which was configured to have 8-bit sampling and
  1024$\times$0.5 MHz channels which more than cover the 320 MHz
  receiver bandwidth. The data are folded on-line with sub-integration
times of 1 min for the AFB and 30 s for the DFB, then written to disk
with 256 bins across the pulse profile for the AFB and 512 bins for
the DFB.  The timing data presented were collected between  January 6,
2000 and March 4, 2014  as detailed in Table \ref{tb1757ob}.

Timing data from the Parkes 64-m radio telescope are also included,
where the observations were carried out between December 2000 and February
2013 with a central observing frequency close to 1.37 GHz. Several
observations of this pulsar were also made with the 10-cm receiver
which has a bandwidth of 1024 MHz centred at 3094 MHz.  An analogue
filterbank and a series of digital filterbanks are used to
acquire the raw data which are openly accessible via the CSIRO Data
Access Portal\footnote{\url{https://data.csiro.au/dap/}}
\citep{hmm+11}. 

The data were reduced using the \textsc{psrchive} pulsar analysis
system \citep{hvm04} to remove radio-frequency interference, and to
dedisperse and average the multi-channel data to form mean pulse
profiles.  The times of arrival (ToAs) of pulses were derived by
cross-correlation with templates or standard profiles which were
obtained by summing all available data for each system.  After that,
the timing analysis was performed with the pulsar timing  package \textsc{tempo2},
\citep{hem06}. The pulse ToAs at the observatories were referred to
the solar system barycenter (SSB) employing the solar system ephemeris
DE421 \citep{fwb08}.  Then the ToAs at the SSB were fitted with the
standard model of the pulse phase, $\phi(t)$, as a function of time,
$t$:
\begin{equation}
\phi=\phi_{\rm 0}+\nu (t-t_{\rm 0})+\frac{1}{2}\dot\nu(t-t_{\rm 0})^2+\frac{1}{6}
\ddot\nu(t-t_{\rm 0})^3
\label{eq:timmod}
\end{equation}
where $\phi_{\rm 0}$ is the phase at time $t_{\rm 0}$ and $\nu$,
$\dot\nu$ and $\ddot \nu$ are the spin frequency, frequency derivative
and frequency second derivative respectively.  A glitch will  result in
an additional pulse phase modelled by the equation \citep{ehm06}:
\begin{equation}\label{eq:gltphi}
  \phi_{\rm g}=\Delta\phi+\Delta\nu_{\rm p}(t-t_{\rm g})+\frac{1}{2}
  \Delta\dot\nu_{\rm p}(t-t_{\rm g})^2+[1-e^{-(t-t_{\rm g})/\tau_{\rm d}}]\Delta\nu_{\rm d}\tau_{\rm d}
\end{equation}
where $t_{\rm g}$ is the glitch epoch and $\Delta\phi$ is an offset of
pulse phase between the pre- and post-glitch data.  The glitch event
is characterised by permanent increments in the spin frequency
$\Delta\nu_{\rm p}$ and first frequency derivative $\Delta\dot\nu_{\rm
  p}$ and a transient frequency increment $\Delta\nu_{\rm d}$ which
decays exponentially with a time scale $\tau_{\rm d}$.  Since ToA
uncertainties are often underestimated because of instrumental and
physical effects, an additional term was added in quadrature to the
ToA uncertainties.  Quoted uncertainties are twice the formal standard
deviation given by \textsc{tempo2} (Tables \ref{tb1757tim} and
  \ref{tb1757fit}).  The offsets between Nanshan AFB data, Nanshan DFB
  data and Parkes data are estimated using \textsc{tempo2}. There are
  no obvious offsets between the different DFB systems at Parkes.
Timing solutions are quoted in the TCB system referenced to
TT(BIPM2013)\footnote{\url{http://www.bipm.org}}.

\section{Analysis and Results}
\subsection{Pulse profiles}

\begin{figure}%[h,t]
\centerline{\includegraphics[angle=-90,width=0.49\textwidth]{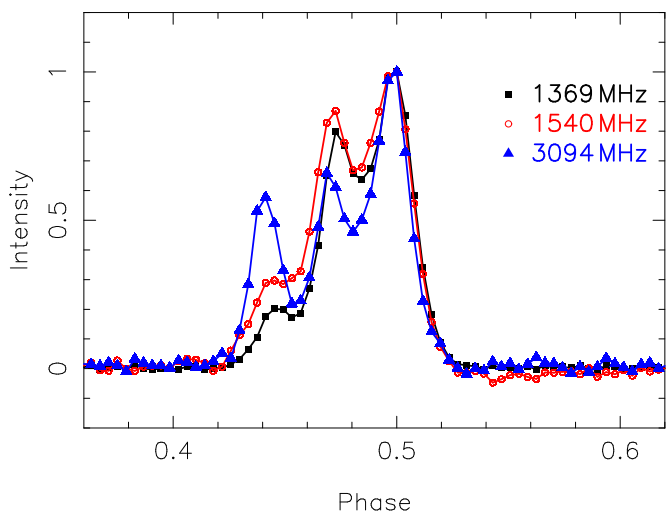}}
\vspace{-2mm}
\caption{The average pulse profiles of PSR J1757$-$2421 at 1369 MHz
  (black), 1540 MHz (red) and 3094 MHz (blue). The pulse peaks are normalised to unity. }
\label{fg:1757prof}
\end{figure}

\begin{figure}%[h,t]
\centerline{\includegraphics[angle=-90,width=0.44\textwidth]{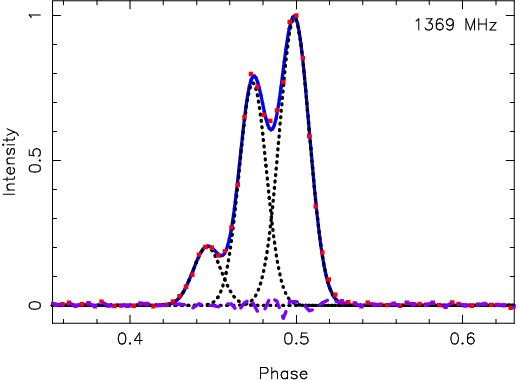}}
\vspace{2mm}
\centerline{\includegraphics[angle=-90,width=0.44\textwidth]{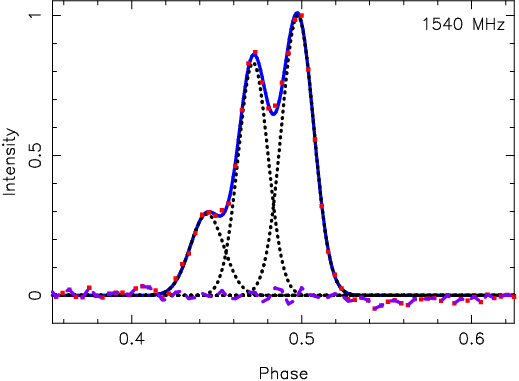}}
\vspace{2mm}
\centerline{\includegraphics[angle=-90,width=0.44\textwidth]{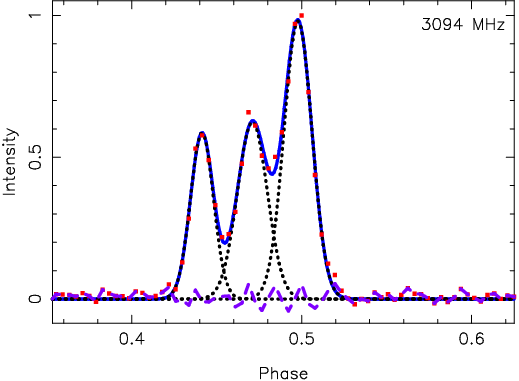}}
\caption{Observed pulse profiles of PSR J1757$-$2421 with fitted
  Gaussian components.  Residuals from the Gaussian fit are shown as a dashed 
 line in each
  plot. }
\label{fg:1757gauss}
\end{figure}

\begin{table*} %[!htb]
\centering
\caption{Parameters of the fitted Gaussian profile components for PSR J1757$-$2421, 
 $I_{\rm i}$ is the amplitude of the $i$th Gaussian component, $\phi_{\rm i}$ is the 
peak phase, and $w_{\rm i}$ is the full width at half peak.}
\label{tb1757gs}
\begin{tabular}{llllllllll}
\hline
Frequency &   $I_1$ & $\phi_1$ & $w_1$ & $I_2$ & $\phi_2$ & $w_2$ & $I_3$ & $\phi_3$ & $w_3$ \\
\hline
1369 MHz  & 0.2019(3) & 0.4463(2)    & 0.0196(3)  & 0.7880(3) & 0.4739(1)    & 0.0193(3)  & 0.9902(3) & 0.49885(6)    & 0.0212(3) \\

1540 MHz  & 0.2913(9) & 0.4443(7)    & 0.0232(10)  & 0.8320(11) & 0.4714(4)   & 0.0205(11)   & 0.9998(10) & 0.4976(2)   & 0.0222(10) \\

3094 MHz   & 0.5847(11)  & 0.4413(2)   & 0.0160(11)  & 0.6236(10) & 0.4710(3)   & 0.0208(10)  & 0.9795(10) & 0.4978(2)   & 0.0193(10)\\
\hline
\end{tabular}
\end{table*}

\begin{figure}%[h,t]
\centerline{\includegraphics[angle=-90,width=0.465\textwidth]{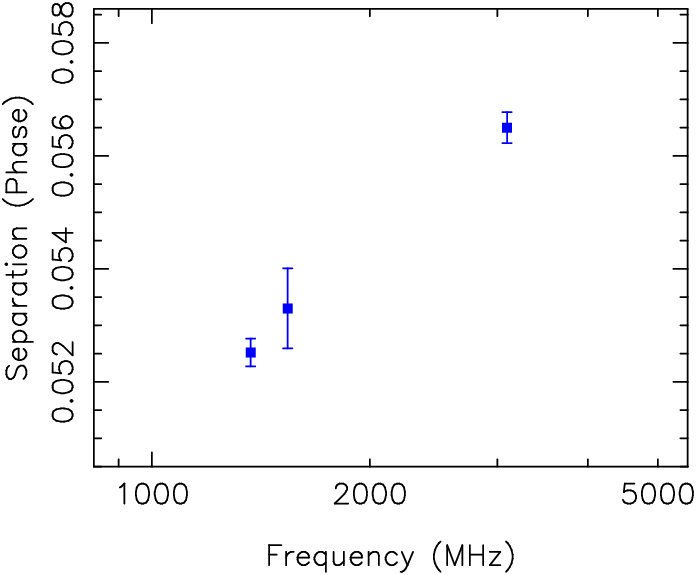}}
\vspace{2mm}
\centerline{\includegraphics[angle=-90,width=0.465\textwidth]{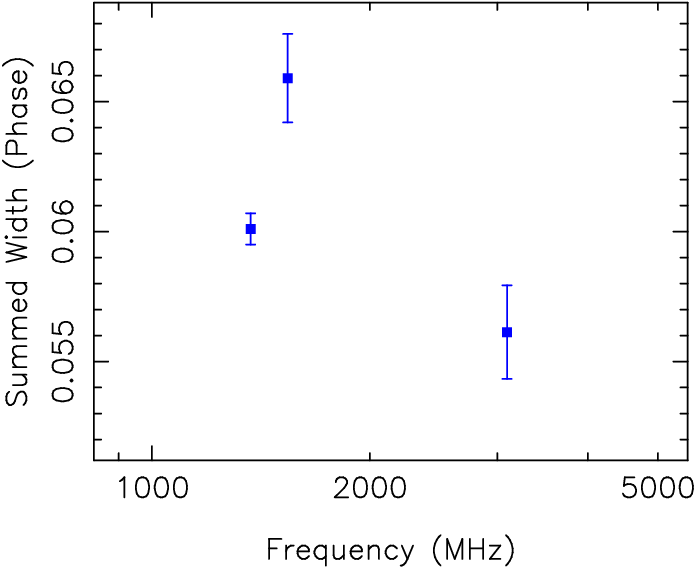}}
\caption{Top: The phase separation of the two outermost components presented in Figure 
\ref{fg:1757gauss} . Bottom: The summed width of the three Gaussian components shown in 
Figure \ref{fg:1757gauss}. }
\label{fg:1757width}
\end{figure}

\begin{figure}%[h,t]
\centerline{\includegraphics[angle=-90,width=0.48\textwidth]{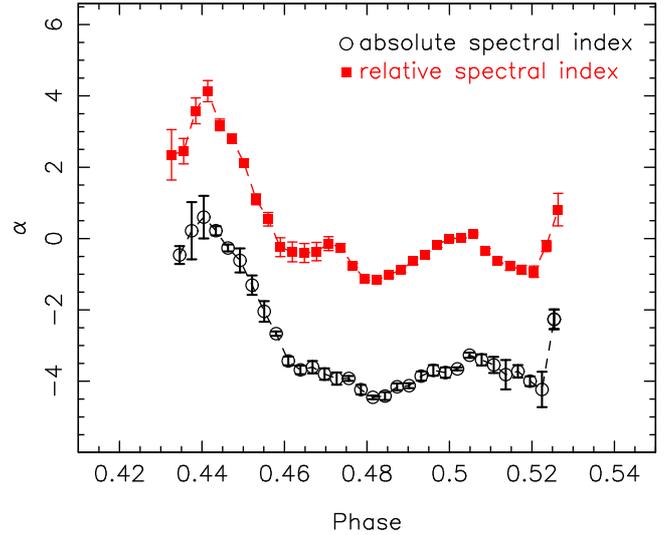}}
\caption{The phase resolved spectral index  for PSR J1757$-$2421.}
\label{fg:1757phspec}
\end{figure}

Average pulse profiles are formed by adding the aligned
pre-glitch data. The profiles at 1369 MHz, 1540 MHz
  and 3094 MHz are shown in Figure \ref{fg:1757prof}, where the pulse
  peaks are normalised to unity. We average the Nanshan DFB and Parkes profiles to 256
  bins to match the Nanshan AFB data.  At all three frequencies the
trailing component is dominant, but the leading component is
relatively much stronger at 3094 MHz.  At 1540 MHz, the width of pulse
at 50\% of peak (w50) is 11.02(14) ms, which implies a duty cycle of
$\sim$ 4.71(6)\%. It is useful to fit the observed profiles by
Gaussian components
 %\citep[e.g.,][]{kwj+94}
 using
\begin{equation} 
I(\phi)=\sum_{i=1}^M I_{\rm i}e^{-4\ln2(\phi-\phi_{\rm i})^2/w_{\rm i}^2}
\end{equation}
where $I_{\rm i}$ is the amplitude of the $i$th Gaussian component,
$\phi_{\rm i}$ is the peak phase, and $w_{\rm i}$ is the full width at
half peak.  These parameters were determined using the Levenberg-Marquardt 
 algorithm \citep{ptvf92}. 
 Three components were fitted
and the fitting results are presented in Table
\ref{tb1757gs} and Figure \ref{fg:1757gauss}. 
It is shown clearly in  Figure \ref{fg:1757width} % 
 that the phase separations of the two outermost components
($\phi_3-\phi_1$) increases with increasing frequency. However,
the summed width of the Gaussian components ($w_1+w_2+w_3$) is smaller at
3.1 GHz.

For most pulsars, the radio spectrum at frequencies above about 1 GHz
obeys a power law $S\sim \nu_r^\alpha$, where $\nu_r$ is the radio
frequency and the observed spectral index $\alpha$ is generally between $-1.5$
and $-1.8$. The relative phase-resolved spectral indices
  shown in Figure \ref{fg:1757phspec} are determined using the
  normalized profiles at 1369 MHz, 1540 MHz and 3094 MHz (Figure
  \ref{fg:1757prof}). The on-pulse region is taken to be where the
  signal exceeds three times the baseline rms noise
  for all frequencies.  As the Nanshan observations were not
  calibrated, we determine absolute spectral indices using the Parkes
  1369 and 3094 MHz data, flux-calibrated using associated
  observations of Hydra A. Figure \ref{fg:1757phspec} shows that the
spectral index for the leading component (phase 0.43 -- 0.45) is close
to zero, whereas the central and trailing components have a much
steeper spectrum with $\alpha \sim -4.0$. The spectral index has its
minimum value between the central and trailing components.

\subsection{The glitch in PSR J1757$-$2421}\label{subg1757}

\begin{figure}%[h,t]
\centerline{\includegraphics[angle=-90,width=0.44\textwidth]{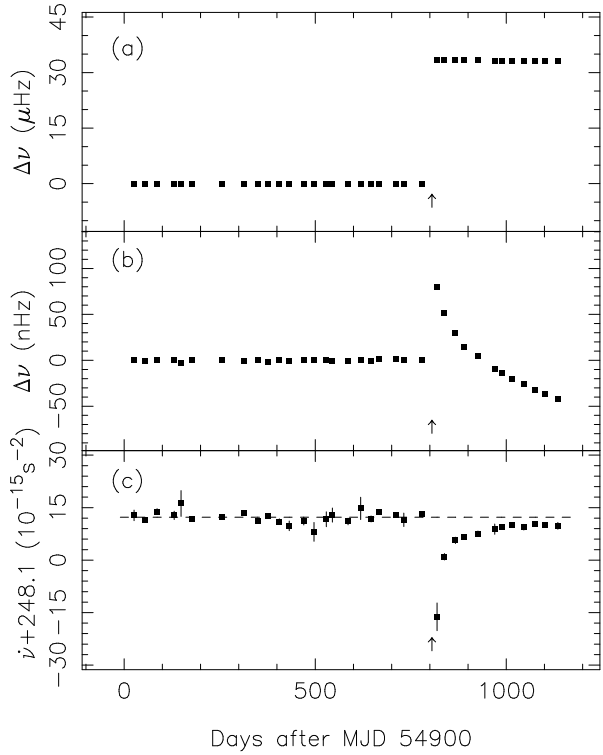}}
\caption{The glitch in PSR J1757$-$2421. The top panel shows the spin-frequency 
residuals $\Delta\nu$ relative to the pre-glitch spin-down solution; the middle panel
 is an expanded plot of $\Delta\nu$ where the mean post-glitch
 value is subtracted  from the post-glitch data; the bottom panel shows the 
variations of $\dot\nu$,  and  the horizontal dashes indicate the average
pre-glitch level.  The glitch epoch is indicated by an 
arrow.}
\label{fg:1757f}
\end{figure}

We adopt the astrometric parameters and dispersion measure obtained by
\citet{hllk05} with the position and proper motion fixed in this
analysis.  It is not possible to obtain a phase-connected solution for
PSR J1757$-$2421 with the whole data set, as a previously unreported
very large glitch occurred in May 2011.  The rotational parameters for
PSR J1757$-$2421 from fitting of the timing model (Equation
\ref{eq:timmod}) to the pre- and post-glitch data are given in Table
\ref{tb1757tim}.  To examine the spin behaviour of PSR
J1757$-$2421, the spin frequency $\nu$ and frequency derivative
$\dot\nu$ were derived from independent fits to short sections of
data, each of which typically span $\sim 60$~d. Figure \ref{fg:1757f}
shows an overview of the spin-down of PSR J1757$-$2421, indicating
that a large glitch with $\Delta\nu \sim 33\times$10$^{-6}$~Hz
occurred between MJD 55697 and 55708 (May 16, 2011 and May 27, 2011).
Assuming a glitch epoch of MJD 55702 and using the data between MJD
54515 and 55795, the spin parameters are extrapolated to the glitch
epoch with the pre- and post-glitch solutions given in Table
\ref{tb1757tim}. It is noted that here we don't take the exponential decays
into account.  Then, the glitch parameters are estimated by
calculating the fractional jump in $\nu$ and $\dot\nu$, which
results in $\Delta\nu_{\rm g}/\nu \sim7809(2)\times10^{-9}$ and
$\Delta\dot\nu_{\rm g}/\dot\nu\sim67(16)\times10^{-3}$.  The frequent timing
observations of this pulsar make it possible to track the recovery
process relatively thoroughly.

\begin{table*}
\centering
\caption{Timing parameters for PSR J1757$-$2421. The uncertainties for the
measured quantities ($\nu$, $\dot\nu$, $\ddot\nu$) are twice the formal 
standard deviations given by \textsc{tempo2} }
\label{tb1757tim}
\begin{tabular}{lll}
\hline%
Parameters & ~Pre-glitch & Post-glitch \\
\hline
Pulsar name\dotfill & ~J1757$-$2421 & J1757$-$2421 \\
MJD range\dotfill & ~51549---55697  & 55707.9---56719.1 \\
Number of ToAs\dotfill & ~407  & 84  \\
Rms timing residual ($\mu s$)\dotfill & ~3203  &  7739 \\
Weighted fit\dotfill &  ~Y  &  Y \\
Reduced $\chi^2$ value \dotfill & ~376.3  &  24454$^{\dag}$ \\
\hline
\multicolumn{3}{c}{Measured Quantities} \\
\hline
Pulse frequency, $\nu$ (s$^{-1}$)\dotfill & ~4.27159264655(4) & 4.2715730980(8) \\
First derivative of pulse frequency, $\dot{\nu}$ (s$^{-2}$)\dotfill & ~$-2.356777(3)\times 10^{-13}$ & $-2.3729(3)\times 10^{-13}$ \\
Second derivative of pulse frequency, $\ddot{\nu}$ (s$^{-3}$)\dotfill & ~$-4.80(13)\times 10^{-25}$& $ 3.4(4)\times 10^{-23}$ \\
\hline
\multicolumn{3}{c}{Set Quantities} \\
\hline
Ecliptic longitude (deg.)\dotfill   & ~269.428041(5) & 269.428041(5) \\
Ecliptic latitude (deg.)\dotfill   & ~$-$0.9307(4) & $-$0.9307(4)\\
Epoch of frequency determination (MJD)\dotfill & ~53623  & 56213 \\
Epoch of position determination (MJD)\dotfill & ~53623  & 53623 \\
Epoch of dispersion measure determination (MJD)\dotfill &  49909  & 49909  \\
Dispersion measure, DM (cm$^{-3}$pc)\dotfill & ~179.454   & 179.454$^\ddag$  \\
Proper motion in ecliptic longitude (mas yr$^{-1}$) & ~$-$17(5) & $-$17(5)\\
\hline
\multicolumn{3}{c}{Assumptions} \\
\hline
Reference time scale \dotfill & ~TT(BIPM2013)  & TT(BIPM2013)  \\
Solar system ephemeris model\dotfill & ~DE421 & DE421 \\
Time units \dotfill  & ~TCB  & TCB  \\
\hline
\multicolumn{3}{l}{$^{\dag}$The reduced $\chi^2$ for post-glitch is large 
because of the unmodelled glitch.} \\
\multicolumn{3}{l}{$^{\ddag}$ The reported variation of DM \citep{hlk+04}
is not considered in this article.}
\end{tabular}
\end{table*}

For this pulsar, the residual frequencies and frequency-derivatives have
been obtained from the values at the various epochs by subtracting the
pre-glitch models. Figure \ref{fg:1757f}(b)  is obtained by the subtraction of
the mean frequency values on each side of the glitch epoch.
Although most of the frequency jump persists beyond the end of the data span,
the expanded plot shown in Figure \ref{fg:1757f}(b) and the plot of $\dot\nu$ 
presented in Figure \ref{fg:1757f}(c) show that there is an initial exponential 
decay. 

\begin{figure}%[h,t]
\centerline{\includegraphics[angle=-90,width=0.44\textwidth]{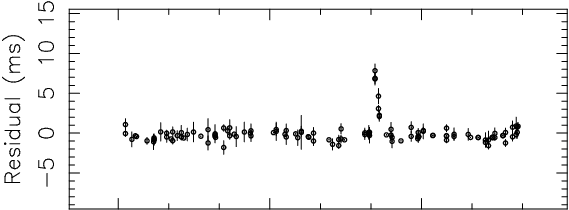}}
\vspace{-1mm}
\centerline{\includegraphics[angle=-90,width=0.44\textwidth]{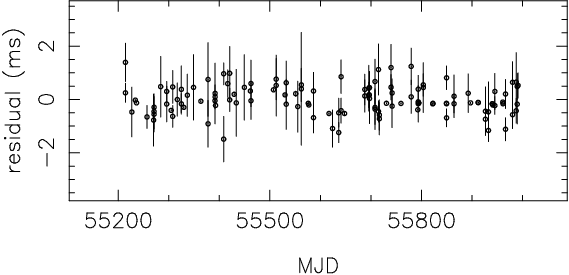}}
\caption{The timing residuals for PSR J1757$-$2421 between MJD 55214 and 55992.
For the upper plot, one decay term is fitted, whereas two exponential decay 
terms are fitted for the lower plot.
}
\label{fg:17571d2d}
\end{figure}

Because of the large amplitude of the glitch, the final result was
obtained in several steps. For the first step, fits are to a data span
that covers 780 d from MJD 55213 to MJD 55993. A single decay with an
assumed time scale of 100 days is fitted to the ToAs, resulting values
of $\Delta\nu_{\rm p}\sim 3.325\times10^{-5}$ Hz.  Post-fit residuals
from the fit are shown in the upper panel of Figure
\ref{fg:17571d2d}. It is obvious from this plot that there is an
additional initial decay with time constant $\sim$15 days. Fitting for
the two decay terms gives $\Delta\nu_{\rm d1}\sim
4.5(23)\times10^{-8}$ Hz and $\tau_{\rm d1}=15(6)$ d, $\Delta\nu_{\rm
  d2}\sim 7.3(8)\times10^{-8}$ Hz and $\tau_{\rm d2}=97(15) $ d.
Residuals for this fit shown in the lower panel of
Figure~\ref{fg:17571d2d} demonstrate that the observed ToAs over this
data span are very well modeled by increments in $\nu$ and $\dot\nu$
together with two exponential decays with time constants around 15 and
100 days, respectively. The resulting glitch parameters are given in
Table~\ref{tb1757fit}. These values are a little larger than those
obtained with the extrapolation method which doesn't take the
exponential recovery into account.  Figure \ref{fg:1757f}(a) shows
that the permanent jump in the frequency is dominant. The fitted
$\Delta\nu_{\rm p}$ =33.263(6) $\times10^{-6}$ Hz, which implies that
99.65\% of the initial frequency jump does not decay. For this event,
the degree of recovery ($Q=\Delta\nu_{\rm d}/\Delta\nu_{\rm g}$),
where $\Delta\nu_{\rm g}=\Delta\nu_{\rm p}+\Delta\nu_{\rm d1}+\Delta
\nu_{\rm d2}$, is very small with $Q\sim0.35(9)$\%. This is consistent
with earlier results that show small values of $Q$, typically around
0.01, are commonly detected in large glitches 
  \citep{ymh+13}.

 The activity parameter for glitching pulsars is defined as
  $A_{\rm g}=\frac{1}{T}\sum\frac{ \Delta\nu_g}{\nu}$, where $T$ is
  the total data span \citep{ml90}. A value of $A_{\rm g}\sim
  0.55\times10^{-9}$~d$^{-1}$ is obtained for PSR J1757$-$2421,
  assuming that it has glitched once since it was discovered in 1974
  and that the 39 years since then is the average glitch interval.
Figure \ref{fg:1757f}(c) shows that there is also a permanent jump in
frequency derivative ($\dot\nu$) of $\dot\nu_{\rm p}\sim
-1.8(4)\times10^{-15} $~s$^{-2}$.  This is small with the fractional increase
$\Delta\dot\nu_{\rm p}/\dot\nu\sim 0.76(9)$\%, less than the value of
1.7\% found by \citet{lsg00} to represent most glitches.  The total
increment in $\dot\nu$ is given by
\begin{equation}
  \Delta\dot\nu_{\rm g}=\Delta\dot\nu_{\rm p} - \Delta\nu_{\rm
    d1}/\tau_{\rm d1} - \Delta\nu_{\rm d2}/\tau_{\rm d2}.
\end{equation}
This gives $\Delta\dot\nu_{\rm g}/\dot\nu=0.196(56)$, much larger than the
permanent increment, but with large uncertainty because of the
uncertainties in the decay parameters. 

\begin{table}%[!htb]

\begin{minipage}{90mm}
\centering
\caption{Glitch parameters for PSR J1757$-$2421. The uncertainties for the
spin parameters are twice the values produced by \textsc{tempo2}}
\label{tb1757fit}
\begin{tabular}{ll}

\hline%\hline
Glitch epoch (MJD)         & 55702 (6) \\
$\Delta \nu_{\rm p}$ (Hz)          &$ 33.263(6)\times10^{-6}$ \\
$\Delta \dot\nu_{\rm p}$ (s$^{-2}$)    & $-1.8(4)\times10^{-15}$ \\

$\Delta\nu_{\rm d1}$  (Hz) & $ 4.5(23)\times10^{-8}$\\
$\tau_{\rm d1}$ (d)   &       15(6)     \\

$\Delta\nu_{\rm d2}$  (Hz) & $7.3(8)\times10^{-8}$ \\ 
$\tau_{\rm d2}$ (d)   &       97(15)     \\

$\Delta\nu_{\rm g}$ (Hz) &   33.381(14)$\times10^{-6}$ \\
$\Delta\nu_{\rm g}/\nu$   &   7815(3) $ \times10^{-9}$ \\
$\Delta\dot\nu_{\rm g}/\dot\nu $ & 196(56) $ \times10^{-3} $\\
$Q (=\Delta\nu_{\rm d}/\Delta \nu_{\rm g})$ & 0.35(9)\% \\
\hline
\end{tabular}
\end{minipage}
\end{table}

\subsection{Timing noise}

\begin{figure}%[h,t]
\vspace{2mm}
\centerline{\includegraphics[angle=-90,width=0.43\textwidth]{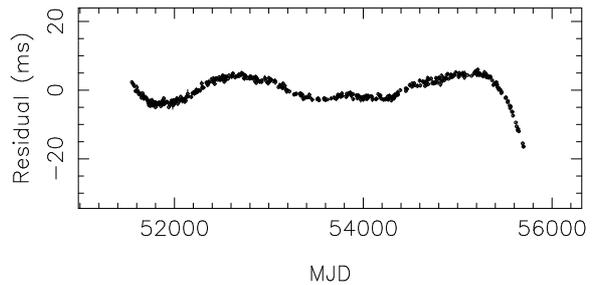}}
\caption{The timing residual of PSR J1757$-2421$ between MJD 51549 -- 55696, with respect to the 
pre-glitch timing model in Table \ref{tb1757tim}.
}
\label{fg:res}
\end{figure}

 Figure~\ref{fg:res} presents the pre-glitch timing residuals relative
 to a solution containing the spin frequency $\nu$, first derivative
 $\dot\nu$ and second derivative $\ddot\nu$ for PSR J1757$-$2421. It
 is obvious that there is significant red timing noise.  The parameter $\Delta_{\rm 8}$ was introduced
 by \citet{antt94} to quantify the magnitude of timing noise. Its
 value is calculated as
\begin{equation}\label{eq:d8}
\Delta_8=\log\left(\frac{|\ddot \nu| }{6\nu} t^3\right)
\end{equation}
where the spin-frequency, $\nu$, and its second derivative,
$\ddot\nu$, are measured over interval of $t=10^8$~s (about 3 yr).  As
the pre-glitch data set for PSR J1757$-$2421 spans over 11 yr, we
obtained four values by fitting over four 3-yr segments, obtaining
$\Delta_8$ ranging from $-$2.33(1) to $-$1.27(3).  It was known that
the $\Delta_{\rm 8}$ parameter is strongly correlated with the
spin-down rate, with $\Delta_8=5.1+0.5\;{\rm log}\dot P$ being
obtained by \cite{hlk10}. For PSR J1757$-$2421 with $\dot
P=12.9\times10^{-15}$ s s$^{-1}$, this relation has a value of $-1.84$,
consistent with the measured values.

\begin{figure}%[h,t]
\centerline{\includegraphics[angle=-90,width=0.46\textwidth]{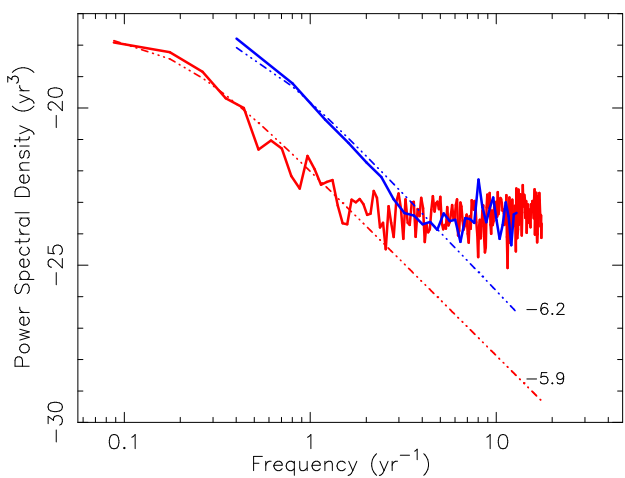}} 
\caption{The observed power spectra  of the pre-glitch timing noise  (red)
 and post-glitch timing noise (blue) for PSR J1757$-2421$. 
The fitted noise models are shown as dashed lines with the spectral exponents presented.}
\label{fg:spc}
\end{figure}

\citet{chc+11} presented a method for analysing correlated timing
noise. This procedure estimates the covariance matrix of timing
residuals using generalized least-squares fitting.  The covariance
matrix is used to perform a linear transformation based on the
Cholesky decomposition of the covariance matrix that whitens both the
residuals and timing model.  Spectral leakage in the presence of
strong red timing noise is minimized with pre-whitening using a
difference filter.  The low-frequency noise can be fitted with a
power-law model $P(f)=A/[1+(f/f_{c})^2]^{\alpha/2}$, where $f$ is the
modulation frequency, $A$ is the amplitude, $\alpha$ is the spectral
exponent, and $f_{c}$ is the corner frequency.  Through an iterative
process, updated estimates of the spectrum can be achieved by using
Lomb-Scargle periodogram after whitening the data using the
Cholesky decomposition of the covariance matrix.  Figure \ref{fg:spc}
presents the spectrum of the pre-glitch timing noise in PSR
J1757$-$2421. The low frequency noise is well modeled by a power-law
with corner frequency of 0.18~yr$^{-1}$ and spectral exponent of
$\alpha\sim -5.9$, which is close to $-6$.  For the post-glitch data,
in order to avoid the impact of the post-glitch recovery, we
analyse the spectrum of the residual between MJD  55822 (100
days after the glitch epoch) and 56720.  Figure \ref{fg:spc} shows the
spectrum of the post-glitch timing noise, which is modeled on a
power-law with spectral exponent of $\alpha\sim -6.2$ and a corner
frequency of 0.4~yr$^{-1}$. Although, the corner frequency and
  spectral exponents are to some degree empirical, these indices suggest that
  both the pre-glitch and post-glitch noise can be modelled
    as a random walk in the spin-down $\dot\nu$.

\begin{table*}
%\scriptsize
\centering
\caption{Parameters for the glitches with two decay terms.}
\label{2day}
\begin{tabular}{lllllllll}
\hline
PSR & Epoch & $\Delta\nu_{d1}$ & $Q_1$ & $\tau_{d1}$ & $\Delta\nu_{d2}$ & $Q_2$ & $\tau_{d2}$ & Ref \\
      &   (MJD) &  ($10^{-6}$ Hz)  &   & (d)  & ($10^{-6}$ Hz) & & (d) \\
\hline J0835$-$4510
              & 40280(4)   &   0.0518(5)     & 0.00198(2)  & 10(1)   &   0.4665(13) & 0.01782(5)   & 120(6)    &\citet{cdk88}  \\
              & 41192(8)   &   0.0362(5)     & 0.00158(2)  & 4(1)    &   0.300(2)   & 0.01311(9    & 94(5)     & \citet{cdk88}  \\
              & 42683(3)   &   0.00968(11)   & 0.000435(5) & 4.0(4)  &   0.0786(4)  & 0.003534(16) & 35(2)     & \citet{cdk88}  \\
              & 43693(12)  &   0.0830(7)     & 0.00242(2)  & 6.0(6)  &   0.3888(7)  & 0.01134(2)   & 75(3)     & \citet{cdk88}  \\
              & 44888.4(4) &   0.01036(10)   & 0.000813(8) & 6.0(6)  &   0.0242(5)  & 0.00190(4)   & 14(2)     & \citet{cdk88}  \\
              & 45192.1(5) &   0.05701(16)   & 0.002483(7) & 3.0(6)  &   0.1263(18) & 0.00550(8)   & 21.5(2.0) & \citet{cdk88}  \\
              & 46259(2)   &   0.066(9)      & 0.0037(5)   & 6.5(5)  &   2.76(1)    & 0.1541(6)    & 332(10)   & \citet{mkhr87} \\
         & 47519.80360(8)  &   0.1086(2)     & 0.005385(10)& 4.62(2) &   3.396(8)   & 0.1684(4)    & 351(1)    & \citet{mhmk90} \\
J1119$-$6127  & 54244(24)  & 18.6(218)  & 0.81(81)   & 15.7(3) &  4.9(43)  & 0.214(136) & 186(3) &  \citet{ymh+13} \\ 
J1757$-$2421  & 55702 (6)   &  0.045(13) & 0.0013(8)  & 15(6) & 0.073(4) & 0.0022(2)  & 97(15) &   this work \\ 
J1803$-$2137  &  50765(15)  &  0.23(16)  & 0.0094(65) & 12(2) &  0.080(15)  & 0.00330(64) & 69(3) &  \citet{ymh+13}\\ 
J2337+6151    &  53615(6)      & 0.19(3) & 0.0046(7)  & 21.4(5) & 0.119(4) & 0.0029(1) & 147(2) &  \citet{ymw+10} \\
\hline\\
\end{tabular}
\end{table*}

\section{Discussion}

The observations presented here show that  the pulse profile exhibits
strong frequency evolution. In numerous early investigations, it was found that the pulse 
component separation often decreases  with increasing frequency
\citep[e.g.][]{tho91}. This is usually interpreted in the context of
the ``magnetic-pole'' model \citep{rc69} as ``radius-to-frequency''
mapping \citep[e.g.,][]{cor78} with the lower frequencies being
emitted at greater altitudes above the magnetic pole. However,
\citet{cw14} found that quite a few pulsars (19\% of 150 normal pulsars) 
showed wider profiles at high frequencies. Figure~\ref{fg:1757width}
shows that PSR J1757$-$2421 is clearly a member of this minority
group.

The average pulse profile of PSR J1757$-$2421 is characterised by
three components whose relative strengths vary with frequency with the
leading component having a much flatter spectral index.  At high
frequencies the conal component, generated from the outer parts of the
open field-line bundle, tends to dominate the profile \citep[e.g.,][]{lm88}. Such kind of
profile evolution may lead to partial cones at certain frequencies
or the broadening of the inner cone width at
high frequency in the inverse Compton scattering model \citep{ql88}.
\citet{wpz+14} account for it as an asymmetrical spectral distribution
across the flux tubes in the pulsar magnetosphere. 

This is the first detection of a glitch in the energetic pulsar PSR
J1757$-$2421 since it was discovered in 1973. This is a Vela-like
event with a relative amplitude of $\Delta\nu_{\rm g}/\nu
\sim7.8\times10^{-6}$, identifying it as large glitch. Such a large
event has not been previously detected in pulsars with characteristic
age $\tau_c \sim 3\times10^5$ yr.  Most of the large
glitches are detected in pulsars with characteristic ages smaller than
10$^5$ yr. For those radio pulsars with characteristic ages 
in the range 2.3$\times10^6$ to 3.2$\times10^6$ yr, similar to PSR J1757$-$2421,
only two (PSRs J1743$-$3150 and J1825$-$0935) have glitched, and these
were small glitches \citep{elsk11}. However, large glitches have been
detected in some pulsars older than PSR J1757$-$2421, such as
J0358+5413 \citep{elsk11}. 

For the glitch in PSR J1757$-$2421, the permanent jump in the spin
frequency is dominant, with $Q\sim 0.35\%$, which is consistent with
the small $Q$ usually seen in large glitches.  PSR J1757$-$2421 shows
an exponential recovery following the glitch with two identifiable
decay terms having time constants of approximately 15~d and 97~d
respectively. Including PSR J1757$-$2421, five pulsars have shown
evidence of two exponential decay terms in one event as listed in Table \ref{2day}.
where the shorter term ranges from 3~d to 22~d, and the longer
range from 14~d to 332~d.  Because observational gaps are often
several weeks, there could be
more fast decays (with time scales less than 20 d) which have been
missed. In this case, the magnitude of the jumps in  $\nu$ and $\dot\nu$ 
could be much larger.

The theoretical understanding of pulsar glitches includes trigger
mechanisms which seek to explain how the glitch originates and to
predict glitch sizes, waiting times and their distributions, and
relaxation mechanisms which explain the behavior and timescales of the
post-glitch recovery.  For large glitches, a popular model is a vortex
unpinning process in the superfluid components of the star
\citep{ai75,aaps84}.  The observed
post-glitch exponential recoveries have been explained as the
re-establishment of an equilibrium between pinning and unpinning in a
vortex-creep region interior to a neutron star \citep{accp93,lsg00}.

\citet{lel99} have defined a coupling parameter, 
 $G = 2\tau_{\rm c} A_{\rm g}$,
 which equals the minimum fraction of the moment of inertia 
that transfers  angular momentum to the crust in glitches.
Since $A_{\rm g}$ for PSR J1757$-$2421 is $\sim 0.55\times 10^{-9}$~d$^{-1}$, the
corresponding $G$ is $\sim 11\%$. \citet{cha05,cha12} found that the
effect of entrainment makes it very difficult to move superfluid
neutrons relative to the crust lattice. As a result, \citet{aghe12}
and \citet{cham13} found that the previous calculations of
\citet{lel99} underestimate by a factor of about four the moment of
inertia required by observed glitches and that the superfluid
reservoir in the crust of neutron stars is insufficient to produce the
observed size and frequency of glitches. For PSR J1757$-$2421, it
requires a reservoir comprising $\sim 45\%$ of the
total moment of inertia.  \citet{dcgf16} calculated the crustal moment
of inertia of glitching pulsars by employing a series of different
unified dense-matter equations of state and found values less than
20\%. Although the computed $G$ for PSR J1757$-$2421 may be an
  over-estimate since only one glitch has been detected, this result
  suggests that transfer of angular momentum from the core superfluid
  may be required to account for the large glitch in this relatively
  old pulsar.

\section*{Acknowledgements}

 This work is supported by  National Basic Research Program of China 
(2015CB857100),  West Light 
Foundation of CAS (No. ZD201302), National Natural Science Foundation of 
China (NSFC No. 11173041) and  Strategic Priority Research Programme (B) 
of  Chinese Academy of Sciences (No. XDB23000000 and XDB09000000). 
JPY acknowledge the FAST Fellowship which is supported by Special Funding 
for Advanced Users, budgeted and administrated by CAMS.
JBW acknowledges support from NSFC Grant (11403086,  U1431107) and West 
light Foundation of CAS (XBBS201322).
XZ is supported by NSFC  (11373006, 11003034).\\

We thank the referee for helpful comments, and H. G. Wang,  J. L. Chen,   K. J. Lee 
and H. Tong for useful discussions.
This work is based on observations made with the Urumqi Nanshan 25 m Telescope, 
which is operated by XAO and the Key Laboratory of Radio Astronomy, Chinese 
Academy of Sciences. The Parkes radio telescope is part of the Australia Telescope, 
which is funded by the Commonwealth of Australia for operation as a National 
Facility managed by the Commonwealth Scientific and Industrial Research Organisation.



%\bibliographystyle{mnras}

\begin{thebibliography}{99}


\bibitem[\protect\citeauthoryear{Alpar et al.}{Alpar et~al.}{1981}]{aaps81}
Alpar M. A., Anderson P. W., Pines D., Shaham J., 1981, ApJ, 249, 29L

\bibitem[\protect\citeauthoryear{Alpar, Anderson, Pines, Shaham}{Alpar et~al.}{1984}]{aaps84}
Alpar M. A., Pines D.,  Anderson P. W., Shaham J., 1984, ApJ, 276, 325

\bibitem[\protect\citeauthoryear{Alpar, Chau, Cheng, Pines}{Alpar et al.}{1993}]{accp93} 
Alpar M. A., Chau H. F., Cheng K. S., Pines, David, 1993, ApJ. 409, 345

\bibitem[\protect\citeauthoryear{Aplpar, Nandkumar \&Pines}{Alpar et~al}{1986}]{anp86}
Alpar M. A., Nandkumar R., Pines D., 1986, ApJ, 311, 197


\bibitem[\protect\citeauthoryear{Andersson, Glampedakis, Ho, Espinoza}
{Andersson et al.}{2012}]{aghe12}
Andersson N., Glampedakis K., Ho W. C. G., Espinoza C. M., 2012, PhRvL, 109, 
241103

\bibitem[\protect\citeauthoryear{Anderson \& Itoh}{Anderson \& Itoh}
{1975}]{ai75}Anderson P. W., Itoh N., 1975, Nature, 256, 25


\bibitem[\protect\citeauthoryear{Arzoumanian, Nice, Taylor \&
  Thorsett}{Arzoumanian et~al.}{1994}]{antt94}
Arzoumanian Z.,  Nice D.~J.,  Taylor J.~H.,    Thorsett S.~E.,  1994, ApJ, 422,
  671

\bibitem[\protect\citeauthoryear{Baym \& Pines}{Baym \& Pines}{1971}]{bp71}
Baym G., Pines D., 1971, Ann. Phys., 66, 816

\bibitem[\protect\citeauthoryear{Boynton et al.}{1972}]{bgh+72}
Boynton P. E., Groth E. J., Hutchinson D. P., Nanos G. P.,
 Partridge R. B., Wilkinson D. T., 1972, ApJ, 175, 217-241

\bibitem[\protect\citeauthoryear{Chamel}{Chamel}{2005}]{cha05}
Chamel N., 2005, Nucl. Phys. A 747, 109 128
 
\bibitem[\protect\citeauthoryear{Chamel}{Chamel}{2012}]{cha12}
 Chamel N., 2012,  Phys. Rev. C 85, 035801 

\bibitem[\protect\citeauthoryear{Chamel}{Chamel}{2013}]{cham13}
Chamel N., 2013, PhRvL, 110a, 1101

\bibitem[\protect\citeauthoryear{{Chen}, {Wang}}{Chen \& Wang}{2014}]{cw14}
Chen J. L., Wang H. G., 2014, ApJS, 215, 11

\bibitem[\protect\citeauthoryear{Cheng}{Cheng}{1987}]{che87}
Cheng K. S., 1987, ApJ, 321, 799

\bibitem[\protect\citeauthoryear{{Coles}, {Hobbs}, {Champion}, {Manchester} \&
  {Verbiest}}{{Coles} et~al.}{2011}]{chc+11}
{Coles} W.,  {Hobbs} G.,  {Champion} D.~J.,  {Manchester} R.~N.,    {Verbiest}
  J.~P.~W.,  2011, MNRAS, 418, 561

\bibitem[\protect\citeauthoryear{Cordes}{Cordes}{1978}]{cor78}
Cordes J. M., 1978, ApJ, 222, 1006

\bibitem[\protect\citeauthoryear{Cordes \& Downs}{Cordes \& Downs}{1985}]{cd85}
Cordes J. M., Downs G. S., 1985, ApJS, 59, 343-382

\bibitem[\protect\citeauthoryear{Cordes, Downs \&  Krause-Polstorff}{Cordes et~al}{1988}]{cdk88}
Cordes J. M., Downs G. S.,  Krause-Polstorff J., 1988, ApJ, 330, 847

\bibitem[\protect\citeauthoryear{ Cordes \& Shannon}{Cordes \& Shannon }{2008}]{cs08}
Cordes J. M. \& Shannon R. M., 2008, ApJ, 682, 1152

\bibitem[\protect\citeauthoryear{Delsate et~al.}{Delsate et~al.}{2016}]{dcgf16}
Delsate T., Chamel N., G\"urlebeck N.,  Fantina A. F.,  Pearson J. M.,  Ducoin C., 2016, 
PhRvD, 94, 023008% preprint (arXiv:1606.00016)

\bibitem[\protect\citeauthoryear{{Edwards}, {Hobbs} \& {Manchester}}
{{Edwards}  et~al.}{2006}]{ehm06}
{Edwards} R.~T.,  {Hobbs} G.~B.,    {Manchester} R.~N.,  2006, \mnras, 372,
  1549


\bibitem[\protect\citeauthoryear{{Espinoza}, {Lyne}, {Stappers} \& {Kramer}}{{Espinoza} et~al.}{2011}]{elsk11}
{Espinoza} C.~M.,  {Lyne} A.~G.,  {Stappers} B.~W.,    {Kramer} M.,  2011,
  MNRAS, 414, 1679

\bibitem[\protect\citeauthoryear{Folkner, Williams \& Boggs}{Folkner  et~al.}{2008}]{fwb08}
Folkner W.~M.,  Williams J.~G.,    Boggs D.~H.,  2008, JPL IOM 343R-08-00


\bibitem[\protect\citeauthoryear{Glampedakis \& Andersson}
{Glampedakis \& Andersson}{2009}]{ga09}
Glampedakis K., Andersson N., 2009, PhRvL, 102, 1101,


\bibitem[\protect\citeauthoryear{Greenstein} {Greenstein}{1970}]{gre70}
Greenstein G., 1970, Nature, 227, 791-794


\bibitem[\protect\citeauthoryear{Haskell \& Antonopoulou} {Haskell \& Antonopoulou} {2014}]
{ha14}Haskell B., Antonopoulou D., 2014, MNRAS, 438, 16

\bibitem[\protect\citeauthoryear{{Hobbs},{Lyne},{Kramer},{Martin} et al}{Hobbs et al.}{2004}]{hlk+04}
Hobbs G., Lyne A. G., Kramer M., Martin C. E., Jordan C., 2004, \mnras, 353, 1311

\iffalse
\bibitem[\protect\citeauthoryear{{Hobbs},{Faulkner} \& {Stairs}} {{Hobbs}
 et~al.}{2004}]{hfs+04}
Hobbs G., et al., 2004 \mnras, 352, 1439
\fi

\bibitem[\protect\citeauthoryear{{Hobbs},{Lorimer},{Lyne},{Kramer}}
{{Hobbs} et~al.}{2005}]{hllk05}
Hobbs G., Lorimer D. R., Lyne A. G., Kramer M., 2005, \mnras, 360, 974

\bibitem[\protect\citeauthoryear{{Hobbs}, {Edwards} \& {Manchester}} {{Hobbs}
  et~al.}{2006}]{hem06}
{Hobbs} G.~B.,  {Edwards} R.~T.,    {Manchester} R.~N.,  2006, \mnras, 369, 655

\bibitem[\protect\citeauthoryear{{Hobbs}, {Lyne} \& {Kramer}}{{Hobbs}
  et~al.}{2010}]{hlk10}
{Hobbs} G.,  {Lyne} A.~G.,    {Kramer} M.,  2010, MNRAS, 402, 1027-1048,

\bibitem[\protect\citeauthoryear{{Hobbs} et al.}{Hobbs et al.}{2011}]{hmm+11}
{Hobbs} G., et al., 2011, PASA, 28, 202


\bibitem[\protect\citeauthoryear{{Hotan}, {van Straten} \&
  {Manchester}}{{Hotan} et~al.}{2004}]{hvm04}
{Hotan} A.~W.,  {van Straten} W.,    {Manchester} R.~N.,  2004, PASA, 21, 302

\bibitem[\protect\citeauthoryear{Jones}{Jones}{1990}]{jon90}
Jones P. B., 1990, MNRAS, 246, 364

\bibitem[\protect\citeauthoryear{Johnston \& Galloway}{Johnston \& Galloway}{1999}]{jg99} 
Johnston S., Galloway D., 1999, MNRAS, 306, 50

\bibitem[\protect\citeauthoryear{Komesaroff}{Komesaroff}{1974}]{kom74}
Komesaroff M. M., 1974, Unpublished

\bibitem[\protect\citeauthoryear{Link, Epstein \& Lattimer}{Link et al.}{1999}]{lel99}
 Link B.,  Epstein R. I.,  Lattimer J. M., 1999, Phys. Rev. Lett. 83, 3362-3365.

\bibitem[\protect\citeauthoryear{Lyne  et~al} {Lyne  et~al}{2010}]{lyn10}Lyne A., Hobbs G., 
Kramer M., Stairs I., Stappers B, 2010, Science, 329, 408

\bibitem[\protect\citeauthoryear{Lyne \& Manchester}{Lyne \& Manchester}
{1988}]{lm88} 
Lyne A. G., Manchester R. N., 1988, \mnras, 234, 477

\bibitem[\protect\citeauthoryear{Lyne, Shemar \& Graham-Simth}{Lyne et~al.}{2000}]
{lsg00}Lyne A. G., Shemar S. L.,  Graham-Smith F., 2000, \mnras, 315, 534

\bibitem[\protect\citeauthoryear{Manchester, Hobbs, Teoh, \&Hobbs}{Manchester et~al}{2005}]{mhth05}
Manchester R. N., Hobbs G. B., Teoh A., Hobbs M., 2005, MNRAS, 129, 1993


\bibitem[\protect\citeauthoryear{McKenna \& Lyne}{McKenna \& Lyne}{1990}]{ml90}
McKenna J.,  Lyne A.~G.,  1990, Nature, 343, 349

\bibitem[\protect\citeauthoryear{McCulloch, Klekociuk, Hamilton,Royle}{McCulloch et al.}{1987}]{mkhr87}
McCulloch P. M., Klekociuk A. R., Hamilton P. A, Royle G. W. R., 1987, AuJPh, 40, 725

\bibitem[\protect\citeauthoryear{McCulloch, Hamilton, McConnell, King}{McCulloch et al.}{1990}]{mhmk90}
McCulloch P. M., Hamilton P. A, McConnell D., King E. A., 1990, Nature, 346, 30


\bibitem[\protect\citeauthoryear{Melatos \& Link }{Melatos \& Link }{2014}]{ml14}
{Melatos} A., {Link} B., 2014, MNRAS, 437, 21-31

\bibitem[\protect\citeauthoryear{{Melatos},{Peralta} \&{Wyite}}
{{Melatos} et~al.}{2008}]{mpw08}
Melatos A., Peralta C., Wyithe J. S. B., 2008, ApJ, 672, 1103


\bibitem[\protect\citeauthoryear{Ou et al.}{Ou et al.}{2016}]
{otkd16}Ou Z. W., Tong H., Kou F. F., Ding G. Q., 2016, MNRAS, 457, 39220

\bibitem[\protect\citeauthoryear{Press et al.}{Press et al.}{1992}]{ptvf92}
Press W. H., Teukolsky S. A., Vetterling V. T., Flannery B. P., 1992, Numerical Recipes in C, 
Cambridge University Press, Cambridge, England 

\bibitem[\protect\citeauthoryear{Qiao \& Lin}{Qiao \& Lin}{1988}]{ql88}
Qiao G. J., Lin W. P., 1998, AA, 333, 172

\bibitem[\protect\citeauthoryear{Radhakrishnan \& Cooke }{Radhakrishnan \& Cooke}
{1969}]{rc69}Radhakrishnan V., Cooke D. J., 1969, ApJ, 3, 225

\bibitem[\protect\citeauthoryear{Radhakrishnan \& Manchester}{Radhakrishnan \& 
Manchester}{1969}]{rm69}Radhakrishnan V., Manchester R. N.,
1969, Nature, 222, 228

\bibitem[\protect\citeauthoryear{{Smith} et al.}{{Smith}
  et~al.}{2008}]{sgc+08} {Smith} D.~A., et al., 2008, A\&A, 492, 923

\bibitem[\protect\citeauthoryear{Thorsett}{Thorsett}{1991}]{tho91}
Thorsett S. E., 1991, ApJ, 377, 263

\bibitem[\protect\citeauthoryear{Urama, Link \& Weisberg}{Urama et~al}{2006}]{ulw06}
Urama J. O., Link B., Weisberg J. M., 2006, MNRAS, 370, 76

\bibitem[\protect\citeauthoryear{Wang et al.}{Wang et~al.}{2001}]{wmz+01}
Wang N.,  Manchester R.~N.,  Zhang J.,  Wu X.~J.,  Yusup A.,  Lyne A.~G.,
  Cheng K.~S.,  Chen M.~Z.,  2001, MNRAS, 328, 855

\bibitem[\protect\citeauthoryear{Wang et al.} {Wang et~al.}{2014}]{wpz+14}
Wang H. G., et al., 2014, ApJ, 789, 73


\bibitem[\protect\citeauthoryear{Yi \& Zhang}{Yi \& Zhang}{2015}]{yz15}
Yi S. X., Zhang S. N., 2015, MNRAS, 454, 3674

\bibitem[\protect\citeauthoryear{Yu et al.}{Yu et~al.}{2013}]{ymh+13}
Yu M., et al., 2013, MNRAS, 429, 688

\bibitem[\protect\citeauthoryear{Yuan et al. }{Yuan et al.}{2010a}]{ywml10}
Yuan J. P., Wang N., Manchester R. N., Liu Z. Y., 2010a, MNRAS, 404, 289

\bibitem[\protect\citeauthoryear{Yuan et al.}{Yuan et al.}{2010b}]{ymw+10}
Yuan J. P., Manchester R. N., Wang N., Zhou X., Liu Z. Y., Gao Z. F., 2010b, ApJL,
719, L111


\bibitem[\protect\citeauthoryear{Zhou et al.}{Zhou et al.}{2014}]{zltx14}
Zhou E. P., Lu J. G., Tong H., Xu R. X. 2014, \mnras, 443, 2705

\end{thebibliography}


%%%%%%%%%%%%%%%%%%%%%%%%%%%%%%%%%%%%%%%%%%%%%%%%%%

%%%%%%%%%%%%%%%%% APPENDICES %%%%%%%%%%%%%%%%%%%%%

%%%%%%%%%%%%%%%%%%%%%%%%%%%%%%%%%%%%%%%%%%%%%%%%%%

% Don't change these lines
\bsp	% typesetting comment

\label{lastpage}
\end{document}